\documentclass[oldversion]{aa}
\usepackage{txfonts}
\usepackage{graphicx}
\usepackage{natbib}
\bibpunct{(}{)}{;}{a}{}{,} 

\begin{document}

\newcommand{\HMS}[3]{$#1^{\mathrm{h}}#2^{\mathrm{m}}#3^{\mathrm{s}}$}
\newcommand{\DMS}[3]{$#1^\circ #2' #3''$}
\newcommand{\TODO}[1]{\textbf{TODO: \emph{#1}}}
\renewcommand{\cite}{\citep} 

\title{Discovery of an X-ray nebula around PSR\,J1718$-$3825 
and implications for the nature of the $\gamma$-ray source HESS\,J1718$-$385}

\titlerunning{PSR\,J1718$-$3825 and HESS\,J1718$-$385}
 
\authorrunning{Hinton et al.}

\author{
J.A.~Hinton \inst{1}
\and S.~Funk \inst{2}
\and S.~Carrigan \inst{3}
\and Y.A.~Gallant \inst{4}
\and O.C.~de~Jager \inst{5}
\and K.~Kosack \inst{3}
\and A.~Lemi\`ere \inst{6}
\and G.~P\"uhlhofer \inst{7}
}

\institute{
School of Physics \& Astronomy, University of Leeds, Leeds LS2 9JT, UK
\and
Kavli Institute for Particle Astrophysics and Cosmology, SLAC, 2575 Sand Hill
Road, Menlo-Park, CA-94025, USA
\and
Max-Planck-Institut f\"ur Kernphysik, P.O. Box 103980, D 69029
Heidelberg, Germany
\and
Laboratoire de Physique Th\'eorique et Astroparticules, CNRS/IN2P3,
Universit\'e Montpellier II, CC 70, Place Eug\`ene Bataillon, F-34095
Montpellier Cedex 5, France
\and
Unit for Space Physics, North-West University, Potchefstroom 2520,
    South Africa
\and
CFA - Harvard, 60 Garden Street, 02138 Cambridge MA, USA
\and
Landessternwarte, Universit\"at Heidelberg, K\"onigstuhl, D 69117 Heidelberg, Germany
}

\date{Received; Accepted}

\abstract{ 
  Combined X-ray synchrotron and inverse-Compton $\gamma$-ray
  observations of pulsar wind nebulae (PWN) may help to elucidate the
  processes of acceleration and energy loss in these systems. In
  particular, such observations provide constraints on the particle
  injection history and the magnetic field strength in these objects.
  The newly discovered TeV $\gamma$-ray source HESS\,J1718$-$385 has
  been proposed as the likely PWN of the high spin-down luminosity
  pulsar PSR\,J1718$-$3825. The absence of previous sensitive X-ray
  measurements of this pulsar, and the unusual energy spectrum of the
  TeV source, motivated observations of this region with
  \emph{XMM-Newton}.
  The data obtained reveal a hard spectrum
 X-ray source at the position of PSR\,1718$-$3825 and evidence for
 diffuse emission in the vicinity of the pulsar. We derive limits on
 the keV emission from the centroid of HESS\,J1718$-$385 and discuss the
 implications of these findings for the PWN nature of this object.

}


\maketitle

\section{Introduction}

Young pulsars drive relativistic winds into their environments,
confinement of which leads to the production of extremely broadband emission via the synchrotron and
inverse-Compton (IC) processes \citep[see][for a recent review]{PWN:review}. 
The most prominent PWN, the Crab
Nebula, is detected in all wavebands from the radio to TeV
$\gamma$-rays~\cite{Whipple:crab}, with the transition from synchrotron
to IC emission 
at $\sim$1~GeV. The recent
increase in sensitivity of ground-based TeV $\gamma$-ray instruments
has led to a rapid increase in the number of putative PWN in this
waveband. These objects are characterised by diffuse, typically
offset, nebulae around high spin-down luminosity pulsars. The
archetype of this new object class is the PWN
G\,18.0---0.7/HESS\,J1825$-$137. G\,18.0---0.7 is a $\sim$5$'$
long asymmetric X-ray synchrotron nebula associated with the
middle-aged (characteristic spin-down age $\tau$=21~kyr) pulsar
PSR\,B1823$-$13~\cite{XMM:1825}.  The IC nebula
HESS\,J1825$-$137 is much larger ($\sim$100$'$ at 1~TeV) but exhibits 
energy-dependent morphology, shrinking towards the pulsar at high
energies~\cite{HESS:1825p2,HESS:1825icrc}, suggestive of cooling of the
highest-energy (X-ray synchrotron emitting) electrons away from the pulsar.

The TeV $\gamma$-ray source HESS\,J1718$-$385 was discovered in deep
observations of the supernova remnant RX\,J1713.7$-$3946 using
H.E.S.S. in 2004-2005~\cite{HESS:twopwn}.  The absence of other
potential counterparts and the relatively compact nature of the source
($9'\times4'$ rms) make an association with PSR\,J1718$-$3825 ($8'$ from
the centroid of the TeV source) plausible.  The TeV source is 
unusual in its sharply peaked spectral energy distribution (SED), which is
similar to that of the $\gamma$-ray nebula of
the Vela pulsar~\cite{HESS:velax}.  The $\gamma$-ray emission from
these objects is commonly attributed to IC scattering of
relativistic electrons \citep[see][for an alternative view]{Horns:VelaX}.  
In this scenario the spectral break seen at $\sim$10~TeV in
these objects can be interpreted as a signature of electron 
cooling. However, PSR\,J1718$-$3825 (estimated distance  4.2~pc)
has a characteristic spin-down age (90~kyr) almost an order of magnitude 
greater than that of the Vela pulsar, making such a high energy break 
very surprising.

The search for a possible X-ray counterpart to HESS\,J1718$-$385 is
important for two reasons: firstly, to verify the identification of
the TeV source as the PWN of PSR\,J1718$-$3825 and secondly, to
explore the physical conditions and electron energy distribution 
in the putative nebula. As no sensitive X-ray observations of the
PSR\,J1718$-$3825/HESS\,J1718$-$385 region existed, 
\emph{XMM-Newton} was used to observe this region in September 2006.

\section{Observations and Analysis}

Observations of HESS\,J1718$-$385 with \emph{XMM-Newton}
were conducted on the $4^{\rm th}$ of September 2006 (Obs.-ID
0401960101). Data of 22.3~ks duration were obtained with all X-ray instruments
(PN, MOS1, MOS2) operating in the full-frame mode with a medium filter.
Our analysis 
utilises the \emph{XMM-Newton} Science Analysis Software (SAS),
version 7.0, together with the Extended Source Analysis Software
package ({\emph{XMM}}-ESAS), version 1.0~\citep{XMM:Snowden}. The 
diffuse-source analysis required the development of our own software
extensions. Following standard calibration and data reduction,
the data were cleaned of 
soft proton flares, reducing the usable observation time to 15.2~ks.


\begin{table}[ht]
\begin{center}
\begin{tabular}{|l|c|c|c|r|} \hline
     ID & XMMU\,J          &   RA           & Dec.         &  \multicolumn{1}{c|}{Counts} \\
	&		   & 17h            & $-38^{\circ}$ &  \\ \hline  
                                                            
      1 & 171813.8$-$382517  &   18$^{\rm m}$13.79$^{\rm s}$ &  25$'$16.6$"$ &  256 \\
      2 & 171808.8$-$382604  &   18$^{\rm m}$08.80$^{\rm s}$ &  26$'$04.0$"$ &  255 \\
      3 & 171833.0$-$382749  &   18$^{\rm m}$33.04$^{\rm s}$ &  27$'$48.9$"$ &  154 \\
      4 & 171830.9$-$382704  &   18$^{\rm m}$30.94$^{\rm s}$ &  27$'$03.7$"$ &  152 \\
      5 & 171824.5$-$383204  &   18$^{\rm m}$24.49$^{\rm s}$ &  32$'$04.2$"$ &  114 \\
      6 & 171805.1$-$383140  &   18$^{\rm m}$05.09$^{\rm s}$ &  31$'$40.5$"$ &  100 \\
      7 & 171723.8$-$383233  &   17$^{\rm m}$23.75$^{\rm s}$ &  32$'$32.8$"$ &   95 \\
      8 & 171803.5$-$383315  &   18$^{\rm m}$03.54$^{\rm s}$ &  33$'$14.6$"$ &   79 \\
      9 & 171853.7$-$382228  &   18$^{\rm m}$53.74$^{\rm s}$ &  22$'$28.0$"$ &   73 \\
     10 & 171847.8$-$384158  &   18$^{\rm m}$47.79$^{\rm s}$ &  41$'$58.2$"$ &   63 \\\hline 
\end{tabular}
\end{center}
\caption{ Properties of the 0.5--10~keV X-ray point sources found in the 
  \emph{XMM-Newton} observation described here, ranked by the number
  of excess counts detected. The statistical errors on the source
  positions lie in the range 0.5--1.4$''$. 
}
\label{tab}
\end{table}

\begin{figure*}
  \hspace{-1.5mm}\resizebox{0.51\hsize}{!}{\includegraphics{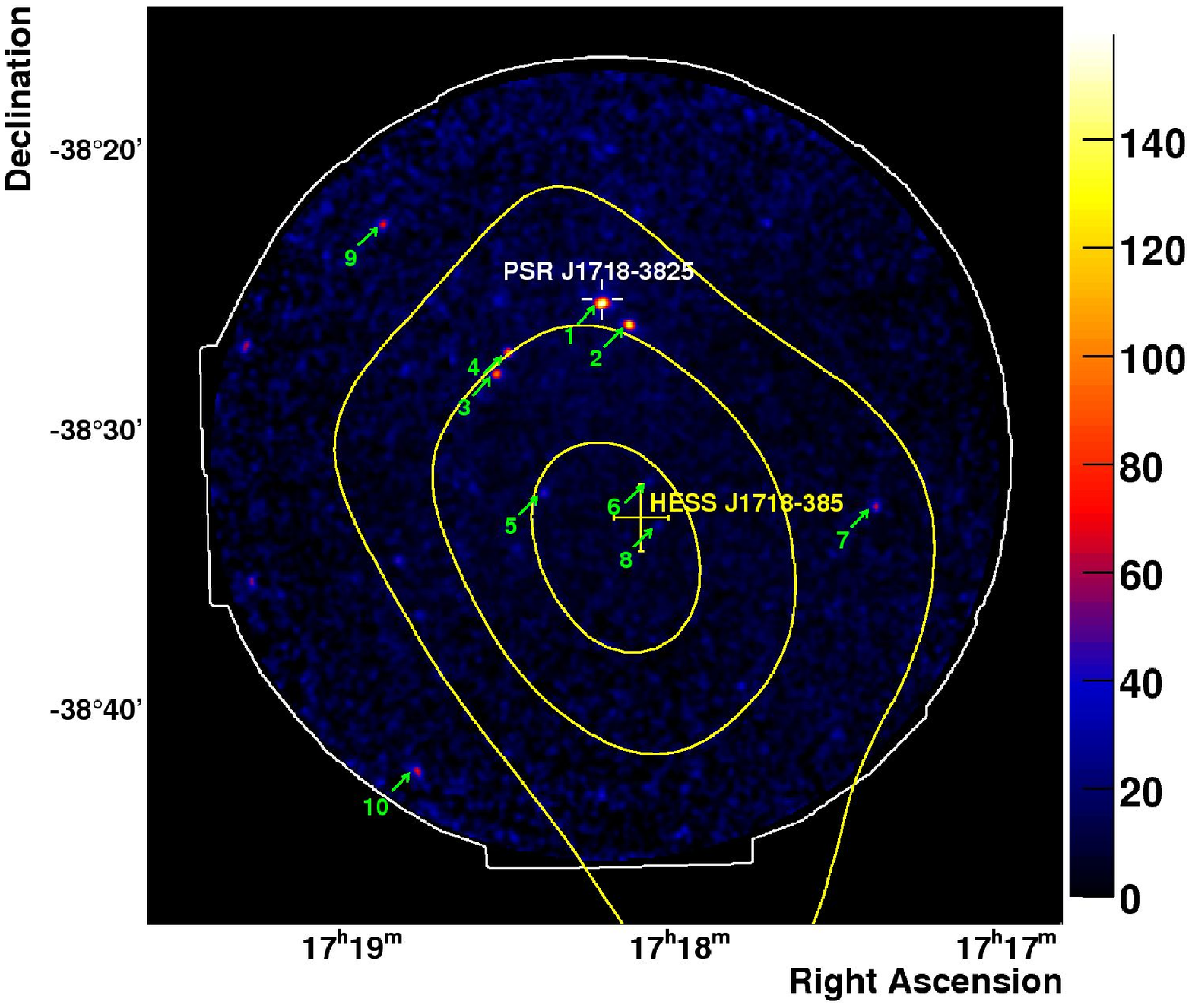}}\hspace{-2mm}\resizebox{0.51\hsize}{!}{\includegraphics{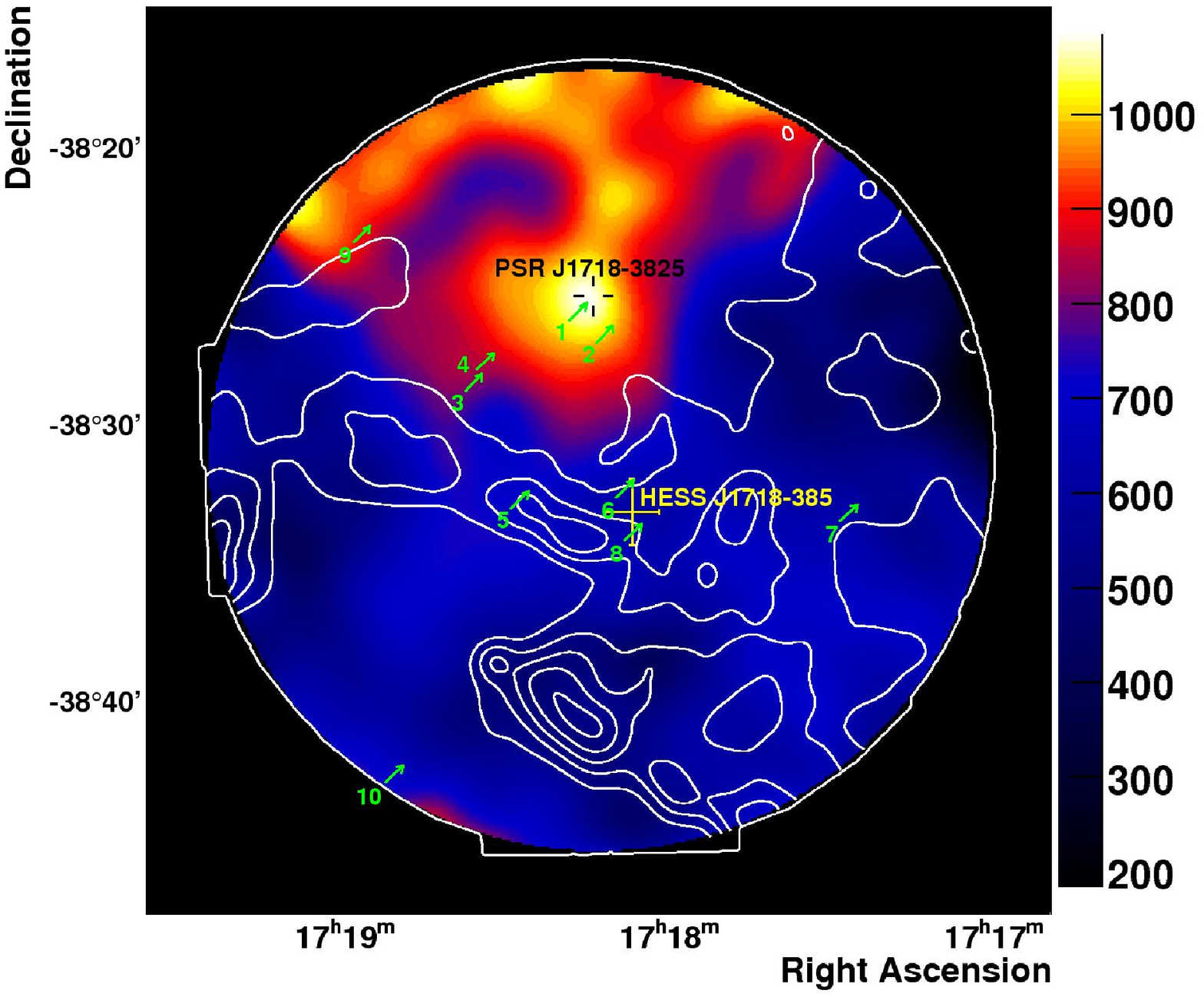}} 
    \caption{
      \emph{Left:} XMM-Newton 0.5--10~keV combined MOS1 and MOS2 image of the vicinity of HESS\,1718$-$385 (colour-scale)
      compared to the TeV $\gamma$-ray morphology (contours). The X-ray data 
      are vignetting-corrected and smoothed with a Gaussian of rms 6$''$.
      The sources summarised in table~\ref{tab} are identified by arrows. 
      The edge of the FoV of the combined MOS1, MOS2 images is
      indicated as a white line. The statistical error on the position of 
      the centroid of HESS\,J1718$-$385 is indicated by a cross. The radio
      position of PSR\,J1718$-$3825 is also marked.
      \emph{Right:} Image of diffuse 0.5--10~keV X-ray emission in the same FoV of
      (colour scale). The \emph{XMM-Newton}
      combined MOS1+MOS2 data have been point-source-subtracted and
      smoothed with a Gaussian of $\sigma=1'$ (see text for
      details). The positions of the subtracted sources 
      are marked with arrows.  The contours show \emph{Spitzer} $8\,\mu$m data from 
      the GLIMPSE survey, smoothed with $\sigma=30''$.
    }	
  \label{fig:maps}
\end{figure*}

Following the standard \emph{XMM-Newton} point-source identification
procedure, ten sources are detected in the combined
MOS1+MOS2 data (see table~\ref{tab}). Each of these sources is
detected in all three standard energy bands (0.5--2~keV,
2--4.5~keV, 4.5--10~keV). The brightest source in the field of
view (source \#1) is coincident with PSR\,J1718$-$3825. 
The other bright sources in the field appear to be associated with
stars: sources \#2,\#3 and \#4 with HD\,323016, HD\,323015 and HD\,323014,
respectively. A smoothed and vignetting-corrected 0.5--10 keV image
is presented in the left panel of Figure~\ref{fig:maps}, with the source fit 
positions marked. Neither of the two previously known X-ray sources in the
field of view (both ROSAT All-sky survey faint sources)
were detected, suggesting that these may be simply statistical
fluctuations (both are $\sim$3$\sigma$ detections) or are perhaps
variable sources. The best-fit position of source \#1 lies 2$"$ away
from the pulsar position. The pointing accuracy of the dataset is
estimated at $\sim$2$"$ from the offsets of sources \#2, \#3 and \#4
from their stellar counterparts, consistent with the specifications of the MOS detectors. 
We therefore conclude that source \#1 is positionally coincident with 
PSR\,J1718$-$3825 and also falls within the region of TeV 
emission HESS\,J1718$-$385. The source position fitting tool {\emph{emldetect}} 
also tries to determine an extension for each source using a
Gaussian emission model. None of the sources is found to be
significantly extended with this tool, however, an
additional spurious source is consistently found to the east of source
\#1 which is not visible upon inspection,
suggesting a possible extension of source \#1.

A point-source spectral analysis of source \#1 using all available data (MOS1, MOS2
and PN) within a radius of 19$"$ yields acceptable goodness-of-fit for
an absorbed power-law spectrum with photon index
$\Gamma=1.47\pm0.21$, a normalisation of $N_0 =
2.4^{+0.9}_{-0.5} \times 10^{-5}$ cm$^{-2}$ s$^{-1}$ keV$^{-1}$ and an
absorbing column $n_{\rm H}$ of $2.9^{+1.4}_{-1.0} \times 10^{21}
\mathrm{cm}^{-2}$ resulting in a flux of $F_{2-10\mathrm{~keV}} \approx
1.4\times 10^{-13}$ erg cm$^{-2}$ s$^{-1}$. 
The best fit $n_{\rm H}$ seems somewhat low considering the
column density of $\approx$2.8$\,\times\,10^{21}$ cm$^{-2}$ in the molecular
component alone (estimated by integrating the $^{12}$CO data of \citet{Dame:12CO}
out to 4.2~kpc). For a typical molecular to atomic ratio (in the range 2--5)
one might expect a value close to $8\,\times\,10^{21}$ cm$^{-2}$. This picture
is also consistent with mean HI column density through the entire galaxy for 
the field of view (FoV) is a whole: $1.2\,\times\,10^{22} \mathrm{cm}^{-2}$.
A fit with an absorbed black-body spectrum does not converge due to the small number of
counts, fitting with a non-absorbed thermal spectrum yields $kT = 1.00\pm0.06$ keV, and a flux of $F_{2-10\mathrm{~keV}} \approx 1.1\times 10^{-13}$ erg
cm$^{-2}$ s$^{-1}$. Fixing $n_{\rm H}$ at $8\,\times\,10^{21}\mathrm{cm}^{-2}$ reduces the best-fit temperature
to $kT = 0.74\pm0.02$ keV. Thermal emission from the neutron star surface
seems very unlikely given the high temperature \citep[see for example][]{Page:neutronstars}
and the emission can be interpreted as non-thermal emission from the pulsar and
an unresolved PWN component. Fixing $n_{\rm H}$ to the same value for 
a power-law fit yields $\Gamma=1.95\pm0.08$. 
For all the spectral fitting, the background was
taken from a ring surrounding the source \#1 (avoiding source \#2) with
inner radius 95$"$ and outer radius 180$"$. 
Given the frame integration times of both the MOS (2~s) and PN (73~ms) cameras
in full frame mode, it was not possible to search for pulsed emission
from source \#1 at the period of PSR\,J1718$-$3825 (74.7~ms). 

To search for diffuse emission, a mask was produced to remove sources
and regions of less than 30\% of the peak exposure in the combined
MOS1+MOS2 data. This mask is applied to both the count map 
(after subtraction of the estimated particle background) and the
exposure map, which are then smoothed and the ratio taken to yield
a vignetting-corrected source-subtracted map (see the right panel 
of Figure~\ref{fig:maps} ). 
There is evidence for diffuse
emission peaking close to PSR\,J1718$-$3825 with a general gradient
in the North-South direction. To test the hypothesis that the non-uniformity
of the large-scale diffuse emission is due to foreground absorption,
we compare the diffuse map to a $8\,\mu$m GLIMPSE image of the 
region---intended to trace dust and hence molecular material along the
line-of-sight. An anti-correlation seems to be present, with the possibility
of substantial absorption towards the centroid of the H.E.S.S. source.
However, such absorption does not seem to be supported by the energy 
dependence of the diffuse emission morphology --- which remains 
roughly constant with energy. Note that the features at the FoV 
edge may be due to under-subtraction of the 
particle background \cite{XMM:Snowden}.
To determine the energy distribution of the diffuse
component around source \#1, a spectrum within an annulus of outer radius 60$"$ and inner 
radius 19$''$ has been determined. A fit of an absorbed
power-law spectrum yields a photon index of
$\Gamma=1.86^{+0.22}_{-0.13}$, 
an absorbing column $n_{\rm H}$ of $7.2^{+3.0}_{-0.8} \times 10^{21}
\mathrm{cm}^{-2}$ and a flux of $F_{2-10\mathrm{~keV}} \approx
1.3\times 10^{-13}$ erg cm$^{-2}$ s$^{-1}$ --- very similar to the total flux
from the central point source \#1. Fixing $n_{\rm H} = 8 \times 10^{21}
\mathrm{cm}^{-2}$ results in $\Gamma=1.84\pm0.09$. There is therefore no
evidence for a spectral softening away from the pulsar, as seen for example
in G\,18.0---0.7, at least not for the inner $1'$. A fit for the full $1'$ radius region yields 
$\Gamma=1.69^{+0.15}_{-0.09}$.

Interpretation of the SED requires measurements
with matching spatial extent. In the absence of a morphological match
between the TeV and diffuse keV emission we can only derive upper limits
on the direct X-ray counterpart to the extended nebula of HESS\,J1718$-$385.
To this end, East-West slices (with a North-South extent matched to the
rms extent 
of the H.E.S.S. source) were made through the vignetting-corrected,
and source-subtracted MOS1+MOS2 count map. These slices are then fit to 
a model of a flat background plus a Gaussian component with an rms width 
constrained to be within a factor 2 of that of the H.E.S.S. source. The 
95\% confidence upper limits on the number of counts in this Gaussian
component are converted into flux upper limits in two energy bands 
(see Figure~\ref{fig:sed}) under the assumption of a $\Gamma=2$ spectrum 
and an absorbing column of $1.2 \times 10^{22} \mathrm{cm}^{-2}$.
As the absorbing column may be significantly higher than this over part
of this region it is important to assess the impact of additional 
absorption on these limits. A factor two increase in the $n_{H}$ assumed
increases the 0.5--2 keV limit by a factor 3, but the 
2--4.5 keV limit by only 30\%.

%



\section{Discussion}

The discovery of hard spectrum X-ray emission from the vicinity of
PSR\,J1718$-$3825, and the evidence for a diffuse halo around
the pulsar strongly suggest the existence of a synchrotron nebula
around this pulsar. This discovery strengthens the association of
the $\gamma$-ray source HESS\,J1718$-$385 to PSR\,J1718$-$3825, but
the relationship of the X-ray emission 
to the $\gamma$-ray source is not straightforward. 
The overall asymmetry of the nebula with respect to the pulsar
is consistent with the idea of SNR expansion into a 
non-uniform molecular environment \citep[see for example][]{blondin01:PWN}.
The very different morphologies in the
two wavebands suggest that either electrons of rather different energies
are responsible for the two sources and/or that the magnetic field
strength within the nebula is highly non-uniform. As the target for
IC emission is the CMBR and other large-scale radiation
fields, the IC flux $F_{\rm IC}$ is simply proportional to the 
number of radiating electrons, $n_{e}$, whereas the 
synchrotron flux goes as: $F_{\rm synch}\propto B^{2}n_{e}$.
In either case the situation may be rather similar to that of HESS\,J1825$-$137 or
indeed HESS\,J1813$-$178 \cite{Funk:1813}, with the lifetime of TeV $\gamma$-ray
emitting electrons being longer than the age of the pulsar, and having time to 
propagate over distances of several parsecs.
The SED of the source is presented in Figure~\ref{fig:sed}.
Three features are of note: 

\begin{figure}[t]
  \centering
  \resizebox{0.9\hsize}{!}{\includegraphics{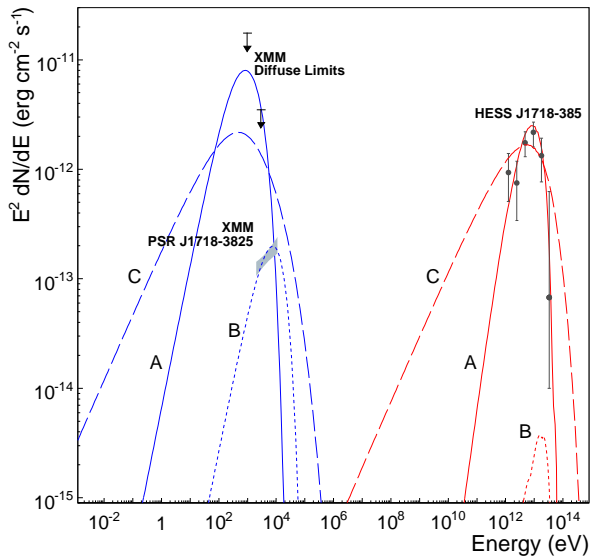}} 
  \vspace{-1mm}
    \caption{Spectral energy distribution for the pulsar wind nebula
      of PSR\,J1718$-$3825. The de-absorbed spectrum of emission from within $1'$
      of the pulsar is shown together with limits for diffuse 
      emission from the region 
      covered by HESS\,J1718$-$385. Three sets of illustrative synchrotron and inverse Compton
      model curves are shown, based on assumptions of: {\bf A)} Mono-energetic 70~TeV electrons injected 
      over a $10^{4}$ year period, $B = 5\mu$G, {\bf B)} 70 TeV electrons, $B=20\,\mu$G, $t=8$ years and {\bf C)} An electron energy distribution 
      following a power-law ($\alpha=1.8$) with an abrupt cut-off at 100~TeV.
      These curves are calculated as described in \citet{GC:Hinton07}.
    }	
      \label{fig:sed}
\end{figure}

1) a hard IC spectrum at TeV energies, with a peak at $\sim$10~TeV.
This suggests that the electrons responsible for
this emission are uncooled. This is rather surprising given the spin-down age of the
pulsar (90 kyr): cooling on the CMBR alone would result in a spectral break at
2~TeV after 90~kyr, inconsistent with the $\gamma$-ray data. Indeed ages $>40$ kyrs 
appear to be excluded by the data. A true age of $\sim$10~kyr could be explained
by a birth period for the pulsar very close to its current period of 75~ms, or 
breaking deviating significantly from the pure magnetic dipole case (as it seems may
commonly be the case, see \citet{Kramer:Spindown})
\footnote{We further note that the projected length of the $\gamma$-ray nebula
($\sim$10~pc) is roughly half that of HESS\,J1825$-$137, despite the fact that
PSR\,J1718$-$3825 is apparently a factor 5 older. Another possibility is that 
 HESS\,J1825$-$137 represents only the youngest part of a larger, softer
spectrum, $\gamma$-ray nebula.}.
The shape of the TeV spectrum appears to be consistent with a constant injection of 
(mono-energetic) $\sim$70~TeV electrons (curve {\bf A}), or with a hard power-law (index $\approx$1.8) 
with a sharp cut-off around 100 TeV ({\bf C}). 

2) hard spectrum X-ray  emission from the pulsar vicinity with a much lower energy flux.
The spectrum from within $1'$ of PSR\,J1718$-$3825 shown in Figure~\ref{fig:sed} 
represents the combined flux of pulsar itself and the inner PWN. The pulsar contribution
is certainly less than half of the total emission (as is clear from the flux in the annulus surrounding the
pulsar) and for typical systems of this type represents only $\sim$20\% of the PWN 
emission in the $>2$keV range \citep{Kargaltsev:ChandraPWN}. Assuming the PWN is 
dominant,
the hard spectrum suggests either higher electron energies or
larger magnetic fields in this region ({\bf B}) in comparison to those found
in the $\gamma$-ray nebula. The lower
flux can be explained if only recently injected electrons are confined
in the region around the pulsar. Whilst the two-zone scenario illustrated by curves {\bf A} and {\bf B} is
clearly grossly oversimplified, 
is does appear that the data are consistent with the idea
that electrons with a relatively 
narrow energy distribution rapidly 
escape from a high $B$-field region close to the pulsar into the extended 
nebula seen in $\gamma$-rays. Another plausible scenario is that the 
injection spectrum of electrons has changed significantly over the
lifetime of the pulsar.

3) the energy flux level of diffuse X-ray emission from the HESS\,J1718$-$385 region
exceeds the TeV flux by not more than a factor $\sim$2. The energy distribution of electrons
is essentially fixed by the TeV data. The diffuse X-ray limits can therefore be used
constrain the $B$-field in the extended  nebula to be not much greater than $5\,\mu$G (curves {\bf C}+{\bf A}), 
close to the mean value of the ISM. We note that this constraint comes principally from the diffuse
limit at 2--4.5 keV which is relatively independent of the assumed
absorbing column.


In conclusion, the diffuse X-ray emission around XMMU\,171813.8$-$382517 
and HESS\,J1718$-$385 appear to represent
different zones in the PWN of the middle-aged pulsar PSR\,J1718$-$3825. Future 
studies of this complex system are certainly well motivated. For example, with
the superior angular resolution of Chandra, the contribution of the pulsar itself
to the non-thermal emission could be separated from that of the PWN.

\bibliographystyle{aa} \bibliography{general}

\end{document}